\documentclass[a4 paper, 11pt]{article}
\pdfoutput=1
\usepackage{comment}
\usepackage{jheppub}
\usepackage{subcaption}
\usepackage{amsmath}
\usepackage{mathrsfs}
\usepackage{amssymb}
\usepackage{amscd}
\usepackage{dsfont}
\usepackage{enumerate}
\usepackage{amsfonts}
\usepackage{epsfig}
\usepackage{mathtools}
\usepackage{yfonts}
\usepackage{bbold}
\usepackage{breqn}
\usepackage[utf8]{inputenc}
\usepackage[english]{babel}
\usepackage{graphicx}
\usepackage{empheq}
\usepackage{relsize}
\usepackage{braket}
\usepackage{caption}
\usepackage{subcaption}

\usepackage{cancel}

\usepackage{jheppub}
\usepackage{mathtools}
\usepackage{float}
\usepackage{tikz-cd}

\date{}
\newcommand{\mc}{\mathcal}

\newcommand{\cd}{\cdots}

\newcommand{\bp}{\bar{\partial}}

 \usepackage[normalem]{ulem}
\usepackage[textsize=scriptsize,textwidth=2.5cm]{todonotes}

\let\exporig\exp
\usepackage{physics}
\usepackage{enumitem}
\let\exp\exporig

\affiliation[\ensuremath{\gamma}]{Yau Mathematical Sciences Center, Tsinghua University, Beijing 100084, China}
\affiliation[\ensuremath{\tau}]{Department of Mathematical Sciences, Tsinghua University, Beijing 100084, China}

\usepackage{xspace}

\allowdisplaybreaks
\newcommand{\z}{\left}
\newcommand{\y}{\right}
\newcommand{\p}{\partial}

\title{
Corrections of an elliptic block in the NS sector
}

\author{Kangning Liu$^{\gamma,\tau}$}

\emailAdd{lkn22@mails.tsinghua.edu.cn}

\abstract
{ 

We propose a correction to one of the elliptic blocks in the NS sector of 2d $\mathcal N = 1$ superconformal field theories. We analyze the 4-point block in the pillow geometry to demonstrate the necessity of the correction and verify the formula by numerically checking the crossing symmetries in the $\mathcal N =1 $ super Liouville theory, as well as directly comparing the $c$-recursion and $h$-recursion results. 

}

\begin{document}
\maketitle
\section{Introduction}
Two-dimensional conformal field theories play a central role in string theory, statistical mechanics, and the study of non‑perturbative aspects of quantum gravity via AdS$_3$/CFT$_2$. Moreover, it is of theoretical interest to consider supersymmetric generalizations of bosonic models. Among them, the \(\mathcal N=1\) super Liouville theory stands out as a prototypical exactly solvable model \cite{Fukuda:2002bv, Nakayama:2004vk}. The correlation functions in $\mathcal N =1$ SCFTs admit a decomposition into superconformal blocks \cite{Belavin:2007gz,Belavin:2007eq,Suchanek:2010kq}. These blocks encode the contribution of a superprimary multiplet exchanged in a given channel and are essential for implementing the conformal bootstrap program\cite{Belavin:1984vu}.

The computation of superconformal blocks has been addressed by two complementary recursive approaches. The \(c\)-recursion formula expresses the block as a sum over its poles in the central charge plane, while the \(h\)-recursion formula performs a similar expansion in the internal conformal dimension. The latter is significantly more efficient in numerical computation, especially when expressed in terms of the elliptic block \(\mathcal{H}(q)\) obtained after factoring out the semi‑classical asymptotics \cite{Zamolodchikov:1987avt}. However, the \(h\)-recursion requires the knowledge of the regular part \(g(q)\) of the elliptic block. For most of the NS sector blocks, this function is known explicitly, except for the case where two of the external fields are descendants of the type $*h = h+\frac{1}{2}$ (i.e. the \(\frac12\) block with two stars, see \ref{sec2}). Earlier work by Hadasz and Suchanek \cite{Hadasz:2007nt} provided a proposal for this regular part, but it was incomplete, and in fact, it depended on all parameters and needed a non-trivial correction.

In this paper, we determine this missing correction up to order \(\mathcal O( q^{15/2})\). Our method combines the exact data obtained from the \(c\)-recursion with the large‑\(h\) asymptotic behavior of the block and the symmetries of the theory. The resulting expression \eqref{correctionfunc} is a polynomial in the external conformal weights whose coefficients are rational functions of the Liouville parameter \(b\) and respect the expected self‑duality \(b\leftrightarrow b^{-1}\). We verify the correctness of our formula through three independent numerical tests: (i) positivity of the coefficients in the pillow geometry for identical external operators, (ii) crossing symmetry of the sphere four‑point function \(\langle VW WV\rangle\), and (iii) direct comparison of the elliptic block computed from the \(c\)-recursion with that obtained from the \(h\)-recursion supplemented by our correction. In all cases, the corrected blocks reproduce the expected results with an error below \(10^{-3}\%\).

The paper is organized as follows. In Section \ref{sec2}, we review the definition and the two recursion formulae of the conformal blocks in the NS sector of $\mathcal N=1$ SCFT; we provide our correction when discussing the $h$-recursion. Section \ref{sec3} contains the numerical checks, including the pillow geometry, crossing symmetry, and direct coefficient comparison. We conclude in Section \ref{sec4} with a summary and outlook.

\section{Conformal blocks in NS sector of $\mathcal N = 1$ SCFT}\label{sec2}

In this section, we will review the 4-point conformal blocks in the NS sector. Let us begin by reviewing the basic setup. 

Consider a unitary 2d SCFT with NS superconformal primary $V_{P}$ with conformal weight $(h_P,\bar h_{P})$, where $P$ is the labelling of the spectrum. There are three other primary fields related to the superconformal primary, denoted as
\begin{equation}
 \Lambda_P =[G_{-\frac{1}{2}},V_P],\quad \bar{\Lambda}_P =[\bar{G}_{-\frac{1}{2}},V_P],\quad W_P=\{G_{-\frac{1}{2}},[\bar{G}_{-\frac{1}{2}},V_P]\}.
\end{equation}
There are two independent three-point function coefficients, defined by the following equations
\begin{equation}
 \begin{aligned}
  &\langle V_{P_1}(z_1)V_{P_2}(z_2)V_{P_3}(z_3)\rangle =\frac{C(P_{1},P_2,P_3)}{|z_{12}|^{2(h_{P_1}+h_{P_2}-h_{P_3})}|z_{13}|^{2(h_{P_1}+h_{P_3}-h_{P_2})}|z_{23}|^{2(h_{P_2}+h_{P_3}-h_{P_1})}}\\
  &\langle W_{P_1}(z_1)V_{P_2}(z_2)V_{P_3}(z_3)\rangle =\frac{\widetilde{C}(P_1,P_2,P_3)}{|z_{12}|^{2(h_1+h_2-h_3+\frac{1}{2})}|z_{13}|^{2(h_1+h_3-h_2+\frac{1}{2})}|z_{23}|^{2(h_2+h_3-h_1-\frac{1}{2})}}
 \end{aligned}
\end{equation}
More specifically, we will consider the $\mathcal N=1$ super Liouville theory, which is parametrized by $b$. Defining $Q=b+b^{-1}$, the central charge is
\begin{equation}
 c=\frac{3}{2}+3Q^2.
\end{equation}
The conformal weights of $V_P$ are
\begin{equation}\label{Theweights}
 h_P=\bar{h}_P= \frac{1}{8}\z(Q^2-4P^2\y) 
\end{equation}
For physical operators, we have $P\in i \mathbb R$\footnote{We do not have to assume unitarity, and for non-unitary super Liouville theories, $P$ can take values on other semi-infinite rays emanating from $P=0$ on the $P$-plane. However, when we discuss the positivity of the four-point block on the pillow, unitarity is necessary. }. The three-point coefficients are functions of $b$ \cite{Rashkov:1996np,Poghossian:1996agj,Fukuda:2002bv, Rangamani:2025wfa, Muhlmann:2025ngz}:
\begin{equation}\label{SC}
 \begin{aligned}
  &C_b(P_1,P_2,P_3)=\frac{\Gamma_b^{NS}(2Q)}{2\Gamma_b^{NS}(Q)^3}\frac{\Gamma^{NS}_b(\frac{Q}{2}\pm P_1\pm P_2\pm P_3)}{\prod_{i=1}^3\Gamma_b^{NS}(Q\pm 2P_i)}\\
  &\widetilde{C}_b(P_1,P_2,P_3)=i\frac{\Gamma_b^{NS}(2Q)}{\Gamma_b^{NS}(Q)^3}\frac{\Gamma^{R}_b(\frac{Q}{2}\pm P_1\pm P_2\pm P_3)}{\prod_{i=1}^3\Gamma_b^{NS}(Q\pm 2P_i)}.
 \end{aligned}
\end{equation}
In the above, the product is over all the permutations of the signs in the numerator. The function $\Gamma^{NS}_b$ and $\Gamma^{R}_b$ are defined as
\begin{equation}
 	\begin{aligned}
		&\Gamma_b^{NS}(x)\equiv\Gamma_b\z(\frac{x}{2}\y)\Gamma_b\z(\frac{x+b+b^{-1}}{2}\y)\\
		&\Gamma^{R}_b(x)\equiv\Gamma_b\z(\frac{x+b}{2}\y)\Gamma_b\z(\frac{x+b^{-1}}{2}\y)
	\end{aligned}
\end{equation}
The double Gamma function is a meromorphic function defined as the unique function satisfying the functional equations
\begin{equation}\label{shiftGamma}
 \begin{aligned}
 \Gamma_b(z+b)&=\frac{\sqrt{2\pi }b^{bz-\frac{1}{2}}}{\Gamma(bz)} \Gamma_b(z)\\
\Gamma_b(z+b^{-1})&=\frac{\sqrt{2\pi }b^{-b^{-1}z+\frac{1}{2}}}{\Gamma(b^{-1}z)} \Gamma_b(z),
 \end{aligned}
\end{equation}
with the normalization given by $\Gamma_b\z(\frac{b+b^{-1}}{2}\y)=1$. For more information about these special functions, see \cite{alexanian2023barnes}, and the appendix of \cite{Collier:2024kwt, Du:2025lya}.

The normalization could be obtained from the above three-point function. Note that
\begin{equation}
  \lim_{P_3\to \frac{Q}{2}}C_b(P_1,P_2,P_3)=\frac{1}{\rho_{NS}^{(b)}(P_1)}(\delta(P_1-P_2)+\delta(P_1+P_2))
\end{equation}
Then the two-point function of two $V_P$'s is:
\begin{equation}
     \langle V_{P_1}(z) V_{P_2}(0)\rangle=\frac{1}{\rho^{(b)}_{NS}(P_1)}\frac{\z(\delta(P_1+P_2)+\delta(P_1-P_2)\y)}{|z|^{2(h_{P_1}+h_{P_2})}}
\end{equation}
where $\rho_{NS}^{(b)}$ is the spectral density in the NS sector:
\begin{equation}\label{measurerho}
 \rho_{NS}^{(b)}(P)=-4\sin\z(\pi bP\y)\sin\z(\pi b^{-1}P\y).
\end{equation}

By inserting the complete basis of states in radial quantization, the spherical four-point functions admit a decomposition using superconformal blocks. Examples include
\begin{align}\label{fourpointfunc}
  &\left\langle V_{P_1}(0) V_{P_2}(z,\bar{z})V_{P_3}(1)V_{P_4}(\infty)\right\rangle_{g=0}\nonumber\\
  &\qquad =\sum_P\Big(C (P_4,P_3,P)C (P,P_2,P_1) \mathcal{F}^1_{h_P}\left[
 \begin{smallmatrix}
  h_{P_3} & h_{P_2} \\
 h_{P_4} & h_{P_1}
 \end{smallmatrix}
  \right](z)\mathcal{F}^1_{h_P}\left[
 \begin{smallmatrix}
  h_{P_3} & h_{P_2} \\
 h_{P_4} & h_{P_1}
 \end{smallmatrix}
  \right](\bar z) \nonumber\\
  &\qquad -\widetilde{C} (P_4,P_3,P)\widetilde{C} (P,P_2,P_1) \mathcal{F}^{\frac{1}{2}}_{h_P}\left[
 \begin{smallmatrix}
  h_{P_3} & h_{P_2} \\
 h_{P_4} & h_{P_1}
 \end{smallmatrix}
  \right](z) \mathcal{F}^{\frac{1}{2}}_{h_P}\left[
 \begin{smallmatrix}
  h_{P_3} & h_{P_2} \\
 h_{P_4} & h_{P_1}
 \end{smallmatrix}
  \right](\bar z) \Big)\nonumber\\
  &\left\langle V_{P_1}(0) W_{P_2}(z,\bar{z})V_{P_3}(1)V_{P_4}(\infty)\right\rangle_{g=0}\nonumber\\
  &\qquad =\sum_P\Big(C (P_4,P_3,P)\widetilde{C} (P,P_2,P_1) \mathcal{F}^1_{h_P}\left[
 \begin{smallmatrix}
  h_{P_3} & *h_{P_2} \\
 h_{P_3} & h_{P_1}
 \end{smallmatrix}
  \right](z) \mathcal{F}^1_{h_P}\left[
 \begin{smallmatrix}
  h_{P_3} & *h_{P_2} \\
 h_{P_3} & h_{P_1}
 \end{smallmatrix}
  \right](\bar z) \nonumber\\
  & \qquad +\widetilde{C} (P_4,P_3,P)C (P,P_2,P_1) \mathcal{F}^{\frac{1}{2}}_{h_P}\left[
 \begin{smallmatrix}
  h_{P_3} & *h_{P_2} \\
 h_{P_4} & h_{P_1}
 \end{smallmatrix}
  \right](z) \mathcal{F}^{\frac{1}{2}}_{h_P}\left[
 \begin{smallmatrix}
  h_{P_3} & *h_{P_2} \\
 h_{P_4} & h_{P_1}
 \end{smallmatrix}
  \right](\bar z) \Big)\nonumber\\
  & \left\langle V_{P_1}(0) W_{P_2}(z,\bar{z})W_{P_3}(1)V_{P_4}(\infty)\right\rangle_{g=0}\nonumber\\
  &\qquad = \sum_P\Big(\widetilde{C} (P_4,P_3,P)\widetilde{C} (P,P_2,P_1) \mathcal{F}^1_{h_P}\left[
 \begin{smallmatrix}
  *h_{P_3} & *h_{P_2} \\
 h_{P_4} & h_{P_1}
 \end{smallmatrix}
  \right](z) \mathcal{F}^1_{h_P}\left[
 \begin{smallmatrix}
  *h_{P_3} & *h_{P_2} \\
 h_{P_4} & h_{P_1}
 \end{smallmatrix}
  \right](\bar z) \nonumber\\
  &\qquad -C (P_4,P_3,P)C (P,P_2,P_1) \mathcal{F}^{\frac{1}{2}}_{h_P}\left[
 \begin{smallmatrix}
  *h_{P_3} & *h_{P_2} \\
 h_{P_4} & h_{P_1}
 \end{smallmatrix}
  \right](z) \mathcal{F}^{\frac{1}{2}}_{h_P}\left[
 \begin{smallmatrix}
  *h_{P_3} & *h_{P_2} \\
 h_{P_4} & h_{P_1}
 \end{smallmatrix}
  \right](\bar z) \Big)\nonumber \\
  &\left\langle V_{P_1}(0) \Lambda_{P_2}(z,\bar{z})\Lambda_{P_3}(1)V_{P_4}(\infty)\right\rangle_{g=0}\nonumber\\
  &\qquad =-\sum_P\Big(\widetilde{C}(P_4,P_3,P)\widetilde{C}(P,P_2,P_1)\mathcal{F}^1_{h_P}\left[
 \begin{smallmatrix}
  *h_{P_3} & *h_{P_2} \\
 h_{P_4} & h_{P_1}
 \end{smallmatrix}
  \right](z)\mathcal{F}^{\frac{1}{2}}_{h_P}\left[
 \begin{smallmatrix}
  h_{P_3} & h_{P_2} \\
 h_{P_4} & h_{P_1}
 \end{smallmatrix}
  \right](\bar{z})\nonumber\\
  &\qquad+C(P_4,P_3,P)C(P,P_2,P_1)\mathcal{F}^{\frac{1}{2}}_{h_P}\left[
 \begin{smallmatrix}
  *h_{P_3} & *h_{P_2} \\
 h_{P_4} & h_{P_1}
 \end{smallmatrix}
  \right](z)\mathcal{F}^{1}_{h_P}\left[
 \begin{smallmatrix}
  h_{P_3} & h_{P_2} \\
 h_{P_4} & h_{P_1}
 \end{smallmatrix}
  \right](\bar{z})\Big)\nonumber\\
 \end{align}
where $*h_i=h_i+\frac{1}{2}$. We also use the symbol $_-h$ to represent either $h$ or $*h$. The signs above arise from exchanging fermionic states.

\subsection{$c$-recursion}
The idea of the $c$-recursion formula is to write the conformal blocks as the sum of poles on the $c$-plane \cite{Belavin:2007gz, Hadasz:2006qb, Balthazar:2022atu}. The locations of these poles are the zeros of the Kac determinant, and are parametrized by two integers $r, s$. Let us introduce the following functions
\begin{align}\label{usefulfunc}
h_{rs} &= - \frac{rs-1}{4} + \frac{1-r^2}{8}b^2 + \frac{1-s^2}{8b^2} \nonumber  \\
\left(b_{r, s}(h)\right)^2 & =-\frac{1}{r^2-1}\left(4 h+r s-1+\sqrt{16 h^2+8(r s-1) h+(r-s)^2}\right) \nonumber\\
c_{r, s}(h) & =\frac{15}{2}+3 b_{r, s}(h)^2+3 b_{r, s}(h)^{-2} \nonumber\\
A_{r, s}(h) & =\frac{1}{2} \prod_{\substack{p=1-r \\
(p, q) \neq(0,0),(r, s);}}^r \prod_{\substack{q=1-s \\ p+q \in 2 \mathbb{Z}}
}^s \frac{\sqrt{2}}{ p b_{r, s}(h)+q b_{r, s}^{-1}(h)} \\
P_{r, s}\left(h_1, h_2\right) & =\prod_{\substack{p \in\{1-r, r-1 ; 2\} \\ q \in \{1-s,s-1;2\} \\
(p+q)-(r+s) \equiv 2 ~ \text{mod} ~4}} \frac{2 P_1-2 P_2-p b_{r, s}-q / b_{r, s}}{2 \sqrt{2}} \frac{2 P_1+2 P_2+p b_{r, s}+q / b_{r, s}}{2 \sqrt{2}}\nonumber\\ 
P_{r, s}\left(h_1, * h_2\right) & =\prod_{\substack{p \in\{1-r, r-1 ; 2\} \\ q \in \{1-s,s-1;2\} \\
(p+q)-(r+s) \equiv 0 ~ \text{mod} ~4}} \frac{2 P_1-2 P_2-p b_{r, s}-q / b_{r, s}}{2 \sqrt{2}} \frac{2 P_1+2 P_2+p b_{r, s}+q / b_{r, s}}{2 \sqrt{2}} \nonumber
\end{align}
Then, one can expand the superconformal blocks as a power series in the cross ratio $z$:
\begin{equation}\label{blockzexpansion}
\begin{aligned}
& \mc{F}^{1}_{h} \z[ \begin{smallmatrix}
  {} _{-} h_3 & {}_{-} h_2 \\
 h_4 & h_1
 \end{smallmatrix} \y](z)=z^{h-h_1- {}_- h_2}\left[1+\sum_{m=\in \mathbb{Z}_{+}} z^m F_m\left(h_4, {}_-h_3, {}_-h_2, h_1 ; h ; c\right)\right] \\
& \mc{F}^{\frac{1}{2}}_{h} \z[ \begin{smallmatrix}
  {} _{-} h_3 & {}_{-} h_2 \\
 h_4 & h_1
 \end{smallmatrix} \y](z)= z^{h-h_1-{}_- h_2} \sum_{k \in \mathbb{Z}_{+}-\frac{1}{2}} z^k F_k\left(h_4, {}_-h_3, {}_-h_2, h_1 ; h ; c\right),
\end{aligned}
\end{equation}
where the coefficients $F_m\left(h_4, {}_-h_3, {}_-h_2, h_1 ; h ; c\right)$ are functions of the external conformal weights $h_i$, the internal conformal weight $h$ and the central charge $c$. As introduced above, these coefficients admit an expansion as a sum over poles in the $c$-plane:
\begin{equation}\label{c-recursion}
 \begin{aligned}
  & F_0 \left(h_4, {}_-h_3, {}_-h_2, h_1 ; h ; c\right) = 1 \\
  & F_m \left(h_4, {}_-h_3, {}_-h_2, h_1 ; h ; c\right) = f_m\left(h_4, {}_-h_3, {}_-h_2, h_1 ; h ; c\right) \\
  &\quad + \sum_{\substack{r=2,3,\cd ,~ s = 1,2,\cd \\ 1<rs\leq 2m,~ r+s \in 2 \mathbb{Z}}} \frac{R^m_{r,s}\left(h_4, {}_-h_3, {}_-h_2, h_1 ; h \right)}{c-c_{r,s}(h)} F_{m-\frac{rs}{2}}\left(h_4, {}_-h_3, {}_-h_2, h_1 ; h+rs/2 ; c_{r,s}(h)\right)
 \end{aligned}
\end{equation}
where the ``seed'' block coefficients $f_m$ appearing on the RHS are the $c\to \infty$ limits of $F_m$, and have the following form
\begin{subequations}
\begin{align}
f_m\left(h_4, h_3, h_2, h_1 ; h\right) & = \begin{cases}\frac{1}{m!} \frac{\left(h+h_3-h_4\right)_m\left(h+h_2-h_1\right)_m}{(2 h)_m} &  m \in \mathbb{Z}_{>0}, \\
\frac{1}{(m-1 / 2)!} \frac{\left(h+h_3-h_4+1 / 2\right)_{m-1 / 2}\left(h+h_2-h_1+1 / 2\right)_{m-1 / 2}}{(2 h)_{m+1 / 2}} &  m \in \mathbb{Z}_{>0}-\frac{1}{2},\end{cases} \\
f_m\left(h_4, h_3, * h_2, h_1 ; h\right) & = \begin{cases}\frac{1}{m!} \frac{\left(h+h_3-h_4\right)_m\left(h+h_2-h_1+1 / 2\right)_m}{(2 h)_m} &  m \in \mathbb{Z}_{>0}, \\
\frac{1}{(m-1 / 2)!} \frac{\left(h+h_3-h_4+1 / 2\right)_{m-1 / 2}\left(h+h_2-h_1\right)_{m+1 / 2}}{(2 h)_{m+1 / 2}} &  m \in \mathbb{Z}_{>0}-\frac{1}{2},\end{cases} \\
f_m\left(h_4, * h_3, h_2, h_1 ; h\right) & = \begin{cases}\frac{1}{m!} \frac{\left(h+h_3-h_4+1 / 2\right)_m\left(h+h_2-h_1\right)_m}{(2 h)_m} &  m \in \mathbb{Z}_{>0}, \\
-\frac{1}{(m-1 / 2)!} \frac{\left(h+h_3-h_4\right)_{m+1 / 2}\left(h+h_2-h_1+1 / 2\right)_{m-1 / 2}}{(2 h)_{m+1 / 2}} &  m \in \mathbb{Z}_{>0}-\frac{1}{2},\end{cases} \\
f_m\left(h_4, * h_3, * h_2, h_1 ; h\right) & = \begin{cases}\frac{1}{m!} \frac{\left(h+h_3-h_4+1 / 2\right)_m\left(h+h_2-h_1+1 / 2\right)_m}{(2 h)_m} & m \in \mathbb{Z}_{>0}, \\
-\frac{1}{(m-1 / 2)!} \frac{\left(h+h_3-h_4\right)_{m+1 / 2}\left(h+h_2-h_1\right)_{m+1 / 2}}{(2 h)_{m+1 / 2}} & m \in \mathbb{Z}_{>0}-\frac{1}{2},\end{cases}
\end{align}
\end{subequations}
where $(\cdot)_m$ stands for the Pochhammer symbol. In addition, the residues $R_{r,s}^m$ of the poles take the following form:
\begin{equation}
\begin{aligned}
     R_{r, s}^m&\left(h_4,{}_- h_3,{}_- h_2, h_1 ; h\right)\\
     &= \begin{cases}\sigma^{r s}\left({}_-h_3\right)\left(-\frac{\partial c_{r, s}(h)}{\partial h}\right) A_{r, s} P_{r, s}\left(h_1,{}_-h_2\right) P_{r, s}\left(h_4,{}_-h_3\right) & \text { for } m \in \mathbb{Z}_{+} \\ \sigma^{r s}\left({}_-h_3\right)\left(-\frac{\partial c_{r, s}(h)}{\partial h}\right) A_{r, s} P_{r, s} (h_1, \widetilde{{}_-h_2} ) P_{r, s} (h_4, \widetilde{{}_-h_3} ) & \text { for } m \in \mathbb{Z}_{+}-\frac{1}{2}\end{cases}
\end{aligned}
\end{equation}
Here, the fusion polynomials $P_{r,s}$ are defined as in \eqref{usefulfunc}. We have also introduced the notation $\sigma^{rs}(*h_3) = (-1)^{rs}$, $\sigma^{rs}(h_3)=1$, and $\widetilde{h}=*h,~ \widetilde{*h}=h$.

\subsection{$h$-recursion}
It is numerically convenient to expand the conformal block as a sum over poles in the $h$-plane. The main difficulty in developing such an expansion is calculating the term regular in $h$. To deal with this problem, Zamolodchikov \cite{Zamolodchikov:1987avt} introduced a semi-classical method to calculate the large $h$ behavior of the Virasoro conformal blocks. Hadasz and Suchanek \cite{Hadasz:2007nt} applied this method to the $\mathcal N = 1$ case. The result is
\begin{equation}\label{largehF}
\begin{aligned}
   & \ln \mc{F}^{1,\frac{1}{2}}_{h} \z[ \begin{smallmatrix}
  {} _{-} h_3 & {}_{-} h_2 \\
 h_4 & h_1
 \end{smallmatrix} \y](z) =  i \pi \tau \z( h - \frac{c-3/2}{24} \y) + \z( \frac{c-3/2}{2} - 4( h_1 - {}_- h_2 -{}_- h_3 - h_4 )\y) \ln \theta_3(q) \\
  &  \qquad + \z( \frac{c-3/2}{24} - {}_-h_2 - {}_-h_3 \y) \ln (1-z) + \z( \frac{c-3/2}{24} - h_1 - {}_-h_2 \y) \ln z +\ln g^{1,\frac{1}{2}}_{{}_{--}}(q) + \mathcal O \z( \frac{1}{h} \y) 
\end{aligned}
\end{equation}
where we introduced the coordinate
\begin{equation}
	\tau(z) = i \frac{K(1-z)}{K(z)}, \quad q = e^{i\pi \tau}, \quad K(z) = {}_2 F_1 \Big( \frac{1}{2}, \frac{1}{2},1 \Big|z \Big)
\end{equation}
and the functions $g^{1,\frac{1}{2}}_{{}_{--}}(q)$ are given by\footnote{Here we only keep the $\mathcal O(h)$ term of $g_{**}^{\frac{1}{2}}(q)$. This is because the $\mathcal{O}(1)$ corrections of $g_{**}^{\frac{1}{2}}(q)$ will only modify the $\mathcal O (1/h)$ term of the conformal block, as in \eqref{fact}. }
\begin{equation}\label{gs}
	\begin{aligned}
		g^1(q ) &= \theta_3(q^2), \qquad & g^{\frac{1}{2}}(q)  & = 0 , \\
		g_*^1(q ) &= \theta_3(q^2), \qquad & g_*^{\frac{1}{2}}(q) & = \theta_2(q^2) , \\
        g_{**}^1(q ) &= \theta_3(q^2), \qquad & g_{**}^{\frac{1}{2}}(q) & = -\theta_2(q^2) h + \mathcal O(1) .
	\end{aligned}
\end{equation}
With this input, one can define the elliptic superconformal blocks \cite{Belavin:2007gz,Hadasz:2007nt,Suchanek:2010kq} as
\begin{equation}\label{FandH}
\begin{aligned}
 \mc{F}^{1,\frac{1}{2}}_{h} \z[ \begin{smallmatrix}
  {} _{-} h_3 & {}_{-} h_2 \\
 h_4 & h_1
 \end{smallmatrix} \y](z) &= (16q)^{h-\frac{c-3/2}{24}} z^{\frac{c-3/2}{24}-h_1-{}_- h_2}(1-z)^{\frac{c-3/2}{24}- {}_- h_2 - {}_- h_3} \\
 &\quad \times \vartheta_3^{\frac{c-3/2}{2}-4(h_1 + {}_- h_2 + {}_- h_3 + h_4)} \mc{H}^{1,\frac{1}{2}}_{h} \z[ \begin{smallmatrix}
  {} _{-} h_3 & {}_{-} h_2 \\
 h_4 & h_1
 \end{smallmatrix} \y](q)
\end{aligned}
\end{equation}
Now, the elliptic block $\mathcal{H}^{1,\frac{1}{2}}_h(q)$ can also be decomposed into a regular function of $h$, i.e., the functions $g^{1,\frac{1}{2}}_{{}_{--}}(q)$, and a sum of poles of $h$. 
Unfortunately, because of the following fact
\begin{equation}\label{fact}
    \ln (c_n h^n + c_{n-1} h^{n-1}+\cd) = \ln (c_n h^n) + \frac{c_{n-1}}{c_n h} + \mathcal O (h^{-2})
\end{equation}
the information of the $\mathcal O(1)$ term of $g^{\frac{1}{2}}_{**}(q)$ is hidden in the $\mathcal O (1/h)$ term in \eqref{largehF}. This is why the determination of $g^{\frac{1}{2}}_{**}(q)$ is subtle. In \cite{Hadasz:2007nt}, the regular function proposed was
\begin{equation}
    \mathfrak{g}^{\frac{1}{2}}_{**}(q) = -\theta_2(q^2) h \z( 1 - \frac{q}{h} \theta_3^{-1} \frac{\p}{\p q} \theta_3(q) + \frac{\theta_2^4(q)}{4h} \y)
\end{equation}
The genuine regular part of the elliptic block should be properly corrected:
\begin{equation}
   {g}^{\frac{1}{2}}_{**}(q) = \mathfrak{g}^{\frac{1}{2}}_{**}(q)+ \mathrm{Correction}_b(h_1,h_2,h_3 , h_4|q)
\end{equation}
As indicated by the form, the final result of ${g}^{\frac{1}{2}}_{**}(q) $ will unfortunately depend on all of the parameters, including $b$ and the external conformal weights $h_i$. Nevertheless, as analyzed in \cite{Hadasz:2007nt}, when $h_i=\frac{1}{8}$ and $c=3/2$, the regular part of the elliptic block $\mc{H}^{\frac{1}{2}}_h (h_1,*h_2,*h_3,h_4)$ is indeed $ \mathfrak{g}^{\frac{1}{2}}_{**}(q)$. Thus, the correction function should satisfy the following constraints:
\begin{itemize}
    \item $\mathrm{Correction}_{b=i}(1/8,1/8,1/8 , 1/8|q)=0$;
    \item $\mathrm{Correction}_b(h_1,h_2,h_3 , h_4|q) = \mathrm{Correction}_{b^{-1}}(h_1,h_2,h_3 , h_4|q)$;
    \item $\mathrm{Correction}_b(h_1,h_2,h_3 , h_4|q) = \mathrm{Correction}_{b}(h_4,h_3,h_2 , h_1|q)$;
    \item $\mathrm{Correction}_b(h_1,h_2,h_3 , h_4|q)$ is at most quadratic in the external conformal weights $h_i$, the central charge $c$ and the inverse $c^{-1}$. 
\end{itemize}
The last constraint comes from the assumption that the classical block limit is well-defined \cite{Zamolodchikov:1987avt}, as the correction function is essentially the $\mathcal O (1/h)$-correction to the classical block. In \cite{Zamolodchikov:1987avt}, Al. Zamolodchikov developed a method to calculate the $\mathcal O(1)$ terms of the classical Virasoro conformal block by inserting a degenerate light operator and analyzing the monodromy of the solutions of the corresponding Fuchsian equation. Twenty years later, the technique was applied in $\mathcal N = 1$ SCFTs in \cite{Hadasz:2007nt}. However, generalizing Zamolodchikov's method to the next order in $\mathcal N=1$ SCFTs is challenging, as all types of four-point functions mix at that order. We will introduce a proposal based on numerical observations in the next subsection that makes the calculation much easier. Although we do not propose a general formula for all orders of the correction, our order $\mathcal O(q^{15/2})$ formula is precise enough for most numerical computations. 

The main result of this paper is
\begin{equation}\label{correctionfunc}
\begin{aligned}
& \mathrm{Correction}_b(h_1,h_2,h_3 , h_4|q)\\
 = & (2h_1-2h_2-2h_3+2h_4) q^{1/2} \\
 & + 8(h_1-h_2-h_3+h_4+2h_1h_3 - 2h_2h_3 - 2h_1h_4 + 2h_2 h_4)q^{3/2} \\
& + C_{5/2} q^{5/2} \\
& + 32(h_1-h_2-h_3+h_4+2h_1h_3 - 2h_2h_3 - 2h_1h_4 + 2h_2 h_4)q^{7/2} \\
& + C_{9/2}q^{9/2} \\
& + 56(h_1-h_2-h_3+h_4+2h_1h_3 - 2h_2h_3 - 2h_1h_4 + 2h_2 h_4)q^{11/2} \\
&  + C_{13/2}q^{13/2} \\
& +96(h_1-h_2-h_3+h_4+2h_1h_3 - 2h_2h_3 - 2h_1h_4 + 2h_2 h_4)q^{15/2} \\
& + \mathcal O (q^{17/2})
\end{aligned}
\end{equation}
where
\begin{subequations}
    \begin{align}
    C_{5/2}  = &  -\frac{9}{8b^4} - \frac{3}{b^2} + \frac{9}{4} - 3b^2 - \frac{9b^4}{8} \nonumber \\
& + \left(20 + \frac{6}{b^2} + 6b^2\right) (h_1 + h_4) + \left(-12 + \frac{6}{b^2} + 6b^2\right) (h_2 + h_3) \\
& - 8 (h_1^2 + h_2^2 + h_3^2 + h_4^2)\nonumber \\
& + 16 h_1 h_2 + 16 h_3 h_4 - 32 ( h_1 h_3 + h_1 h_4 + h_2 h_3 + h_2 h_4 ) \nonumber \\
  C_{9/2}  = &   \frac{43}{4} - \frac{27}{8b^4} - \frac{9}{b^2} - 9b^2 - \frac{27b^4}{8}\nonumber \\
& + \left(54 + \frac{18}{b^2} + 18b^2\right) (h_1 + h_4) + \left(-46 + \frac{18}{b^2} + 18b^2\right) (h_2 + h_3) \\
& - 24 (h_1^2 + h_2^2 + h_3^2 + h_4^2) \nonumber \\
 & - 16 h_1 h_2 - 16 h_3 h_4 - 64 (h_1 h_3 + h_1 h_4 + h_2 h_3 + h_2 h_4)\nonumber \\
  C_{13/2}  = & \frac{45}{4} - \frac{45}{8b^4} - \frac{15}{b^2} - 15b^2 - \frac{45b^4}{8} \nonumber \\
& + \left(100 + \frac{30}{b^2} + 30b^2\right) (h_1 + h_4) + \left(-60 + \frac{30}{b^2} + 30b^2\right) (h_2 + h_3) \\
& - 40 (h_1^2 + h_2^2 + h_3^2 + h_4^2) \nonumber \\
&+ 80 h_1 h_2 + 80 h_3 h_4 - 160 (h_1 h_3 + h_1 h_4 + h_2 h_3 + h_2 h_4)  \nonumber
\end{align}
\end{subequations}
One can check that the correction satisfies all the constraints listed in the previous paragraph. Once the regular parts $g^{1,\frac{1}{2}}_{{}_{--}}(q)$ of $\mathcal{H}^{1,\frac{1}{2}}_h(q)$ are determined, one can write down the $h$-recursion formulae
\begin{equation}\label{Hrec}
    \begin{aligned}
       & \mc{H}^{1}_{h} \z[ \begin{smallmatrix}
  {} _{-} h_3 & {}_{-} h_2 \\
 h_4 & h_1
 \end{smallmatrix} \y](q) = g^1_{{}_{--}}(q) \\
   &\qquad  + \sum_{r,s \in 2\mathbb{Z}_{>0}} (16 q)^{\frac{rs}{2}} \frac{A_{r,s} P_{r,s}(h_1,{}_-h_2)P_{r,s}(h_4,{}_-h_3)}{h-h_{rs}} \mc{H}^{1}_{h_{rs}+\frac{rs}{2}} \z[ \begin{smallmatrix}
  {} _{-} h_3 & {}_{-} h_2 \\
 h_4 & h_1
 \end{smallmatrix} \y](q) \\
  & \qquad + \sum_{r,s \in 2\mathbb{Z}_{\geq 0}+1} (16 q)^{\frac{rs}{2}} \frac{\sigma^{rs}({}_-h_3) A_{r,s} P_{r,s}(h_1,{}_-h_2)P_{r,s}(h_4,{}_-h_3)}{h-h_{rs}} \mc{H}^{\frac{1}{2}}_{h_{rs}+\frac{rs}{2}} \z[ \begin{smallmatrix}
  {} _{-} h_3 & {}_{-} h_2 \\
 h_4 & h_1
 \end{smallmatrix} \y](q) \\
 & \mc{H}^{\frac{1}{2}}_{h} \z[ \begin{smallmatrix}
  {} _{-} h_3 & {}_{-} h_2 \\
 h_4 & h_1
 \end{smallmatrix} \y](q) =g^{\frac{1}{2}}_{{}_{--}}(q) \\
   & \qquad + \sum_{r,s \in 2\mathbb{Z}_{>0}} (16 q)^{\frac{rs}{2}} \frac{A_{r,s} P_{r,s}(h_1,\widetilde{{}_-h_2})P_{r,s}(h_4,\widetilde{{}_-h_3})}{h-h_{rs}} \mc{H}^{\frac{1}{2}}_{h_{rs}+\frac{rs}{2}} \z[ \begin{smallmatrix}
  {} _{-} h_3 & {}_{-} h_2 \\
 h_4 & h_1
 \end{smallmatrix} \y](q) \\
  & \qquad + \sum_{r,s \in 2\mathbb{Z}_{\geq 0}+1} (16 q)^{\frac{rs}{2}} \frac{\sigma^{rs}({}_-h_3) A_{r,s} P_{r,s}(h_1,\widetilde{{}_-h_2})P_{r,s}(h_4,\widetilde{{}_-h_3})}{h-h_{rs}} \mc{H}^{1}_{h_{rs}+\frac{rs}{2}} \z[ \begin{smallmatrix}
  {} _{-} h_3 & {}_{-} h_2 \\
 h_4 & h_1
 \end{smallmatrix} \y](q)
    \end{aligned}
\end{equation}
Numerically, $h$-recursion for elliptic blocks $\mathcal{H}_h^{1,\frac{1}{2}}$ is much more efficient than the $c$-recursion for $\mathcal{F}_h^{1,\frac{1}{2}}$. If one uses \eqref{FandH} to calculate $\mathcal{H}_h^{1,\frac{1}{2}}$ from the $c$-recursion of  $\mathcal{F}_h^{1,\frac{1}{2}}$, the size of the formula will be about twenty times that of directly using \eqref{Hrec}. Although the two formulae should be essentially identical, it is numerically inefficient to simplify the formula of the $c$-recursion one.

\subsection{Determination of the correction}
Firstly, one can use the definition \eqref{FandH} to directly calculate the elliptic blocks. That is, one can use the $c$-recursion formula \eqref{c-recursion} to generate $\mathcal F^{1,\frac{1}{2}}_h$ first, and then use \eqref{FandH} to calculate $\mathcal H^{1,\frac{1}{2}}_h$. 
Indeed, using this method, we can obtain all the corrections up to order $\mathcal O(q^{11/2})$ by using, e.g., Mathematica. However, for the order $\mathcal O(q^{13/2})$ and so on, this method is numerically inefficient. To generate the correction more efficiently, further insights are needed. 

One possible observation is that, since the data we need is nothing but the $\mathcal O(1/h)$-term in the expansion of $\ln \mc{F}^{\frac{1}{2}}_h$ in \eqref{largehF}, maybe there is some relation between the $\mathcal O(1/h)$-terms of different $\ln \mc{F}^{\frac{1}{2}}_h$'s. Based on various numerical checks, our main proposal is: 
\begin{equation}\label{mainprop}
\begin{aligned}
    &\text{The $\mathcal O(1/h)$-term of } \z(  \ln \mc{F}^{\frac{1}{2}}_{h} \z[ \begin{smallmatrix}
 *  h_3 & * h_2 \\
 h_4 & h_1
 \end{smallmatrix} \y](z) - \ln \mc{F}^{\frac{1}{2}}_{h} \z[ \begin{smallmatrix}
   h_3 + \frac{1}{2} & * h_2 \\
 h_4 & h_1
 \end{smallmatrix} \y](z) \y) \\
 & \qquad \qquad \qquad  \qquad \qquad \qquad \qquad \qquad \qquad \qquad \qquad = \frac{A(q) h_3 + B(q) h_4}{h}
\end{aligned}
\end{equation}
where $A(q),B(q)$ are functions of $q$ solely, independent of $h_i$ and $b$. Note that the notation $h_3 + \frac{1}{2}$ means to simply shift the conformal weight $h_3$ to be $h_3 + \frac{1}{2}$ rather than acting a supercharge on $h_3$; the latter should be denoted as $* h_3$. A proof of this proposal would be desirable, but the author has not yet found one.

This proposal is useful, since the $\mathcal O(1/h)$-terms of $\ln \mc F^{\frac{1}{2}}_h$'s are exactly the $\mathcal O(1/h)$-terms of $\ln \mc H^{\frac{1}{2}}_h$'s. And the $\mathcal O(1/h)$-term of $\ln \mc H^{\frac{1}{2}}_h(h_1,*h_2,h_3+1/2,h_4)$ can be easily calculated using the $h$-recursion formula, since $g^{\frac{1}{2}}_*(q)$ is already known in \eqref{gs}. Thus the $\mathcal O(1/h)$-term of $\ln \mc H^{\frac{1}{2}}_h(h_1,*h_2,*h_3,h_4)$ is fixed up to two unknown $q$-dependent coefficients in front of $h_3$ and $h_4$. These two coefficients can be determined using the symmetry
\begin{equation}\label{swapsym}
    \mathrm{Correction}_b(h_1,h_2,h_3 , h_4;h|q) = \mathrm{Correction}_{b}(h_4,h_3,h_2 , h_1;h|q)
\end{equation}
For instance, we have
\begin{equation}
	\begin{aligned}
		& \text{The $\mathcal O(1/h)$-term of } \z( \ln \mc{H}^{\frac{1}{2}}_{h} \z[ \begin{smallmatrix}
   h_3 + \frac{1}{2} & * h_2 \\
 h_4 & h_1
 \end{smallmatrix} \y](q) \y)\\
  & \qquad \qquad \qquad \qquad  \ni \frac{q^{5/2}}{h} \Big( 6 h_2 -2 h_3 + \text{terms satisfying the symmetry in \eqref{swapsym}} \Big)
	\end{aligned}
\end{equation}
Since the difference can only depend on $h_3$, we claim that
\begin{equation}
	\begin{aligned}
		& \text{The $\mathcal O(1/h)$-term of } \z( \ln \mc{H}^{\frac{1}{2}}_{h} \z[ \begin{smallmatrix}
   * h_3  & * h_2 \\
 h_4 & h_1
 \end{smallmatrix} \y](q) \y)\\
  & \qquad \qquad \qquad \qquad  \ni \frac{q^{5/2}}{h} \Big( 6 h_2 + 6 h_3 + \text{terms satisfying the symmetry in \eqref{swapsym}} \Big)
	\end{aligned}
\end{equation}

\section{Numerical checks}\label{sec3}
In this section, we provide numerical checks from various aspects. 

\subsection{Pillow correlation function}
Let us briefly review the idea of the pillow geometry\footnote{We thank Chi-Ming Chang for helpful discussion regarding this subsection. }. For more discussion regarding the physical aspects, see \cite{Maldacena:2015iua}. The variable $q=e^{i\pi \tau(z)}$ has a geometric interpretation: $\tau$ is the modulus of the torus described by the equation
\begin{equation}
    y^2 = x(z-x)(1-x)
\end{equation}
There is a $\mathbb{Z}_2$ symmetry $y\mapsto -y$. The quotient $T^2/\mathbb{Z}^2$ is a Riemann sphere which is flat except for four conical defects at the four fixed-points of $\mathbb{Z}_2$. It is helpful to introduce a coordinate $u$ by
\begin{equation}
    du = \frac{1}{\theta_3 (q)^2} \frac{dx}{y}
\end{equation}
Then the $\mathbb{Z}_2$ acts as $u\mapsto -u$ and the four fixed points are $u_1=0, ~u_2=\pi,~u_3 = \pi(\tau+1),~u_4 = \pi \tau$. The pillow geometry can be transformed into $\mathbb{C}P^1$ via a Weyl transformation, which will introduce a Weyl factor on the four-point function. And the four-point functions in the two geometries differ only by this factor:
\begin{equation}\label{CPPillow}
	\begin{gathered}
		\langle V_1 \cd V_4 \rangle_{\mathbb{C}P^1}(z) = \Lambda(z)\bar{\Lambda}(\bar z) \langle V_1(u=0) V_2(u=\pi) V_3(u=\pi(1+\tau)) V_4(u=\pi \tau) \rangle_{\mathrm{pillow}} \\
		 \Lambda(z) = \theta_3(q)^{\frac{c}{2}-4(h_1+\cd +h_4)} z^{\frac{c}{24}-h_1-h_2} (1-z)^{\frac{c}{24}-h_2-h_3}
	\end{gathered}
\end{equation}
The cycles of the original torus $T^2$ now become cycles of the pillow. We can choose a cycle $A$ that separating pairs of conical defects $u=0,\pi$ from $u=\pi \tau , \pi (1+\tau)$. Then, we can quantize the theory by choosing $A$ as the spatial slice. In this way, the correlation function on the pillow can be understood as a sum over states:
\begin{equation}
	\begin{gathered}
		\langle V_1(u=0) V_2(u=\pi) V_3(u=\pi(1+\tau)) V_4(u=\pi \tau) \rangle_{\mathrm{pillow}}  = \bra{\psi'} q^{L_0 - \frac{c}{24}} \bar{q}^{\bar{L}_0 - \frac{c}{24}} \ket{\psi}\\
		\ket{\psi} \equiv \ket{V_3(u=\pi(1+\tau)) V_4(u=\pi \tau)}_{\mathrm{pillow}} \\
		\ket{\psi'} \equiv \ket{V_1(u=0) V_2(u=\pi )}_{\mathrm{pillow}}
	\end{gathered}
\end{equation}
Now, if $V_4=V_1$ and $V_3=V_2$, the pillow correlation function becomes a sum of inner products
\begin{equation}
\begin{gathered}
    \ket{\psi} = \ket{\psi'} = \sum_{n} c_n \ket{E_{L,n},E_{R,n}} \\ 
	\langle V_1(u=0) V_2(u=\pi) V_2(u=\pi(1+\tau)) V_1(u=\pi \tau) \rangle_{\mathrm{pillow}} = \sum_{n} |c_n|^2 q^{E_{L,n}} \bar{q}^{E_{R,n}}
\end{gathered}
\end{equation}
In particular, the coefficients of the $q$-expansion are positive. One can introduce conformal blocks for the 4-point functions on the pillow
\begin{equation}
	\langle V_1(u=0) V_2(u=\pi) V_3(u=\pi(1+\tau)) V_4(u=\pi \tau) \rangle_{\mathrm{pillow}} = \sum_{P} f_{1,2,P} f_{P,3,4} \widetilde{\mathcal{V}} (q) \widetilde{\mathcal{V}}_{\bar h}(\bar q)
\end{equation}
Using \eqref{CPPillow}, for $\mathcal N =1$ SCFTs, we have
\begin{equation}
	\widetilde{\mathcal{V}}(q) = (16q)^{h(P)-\frac{c}{24}} \prod_{n=1}^{\infty} \z( 1 - q^{2n} \y)^{-\frac{3}{4}} \mathcal{H} (q)
\end{equation}
For our purpose, we should consider
\begin{equation}\label{VHellip}
\begin{aligned}
 	&\widetilde{\mc{V}}^{\frac{1}{2}}_{h} \z[ \begin{smallmatrix}
  * h_3 & * h_2 \\
 h_4 & h_1
 \end{smallmatrix} \y](q) =(16q)^{h(P)-\frac{c}{24}} \prod_{n=1}^{\infty} \z( 1 - q^{2n} \y)^{-\frac{3}{4}}  {\mc H}^{\frac{1}{2}}_{h} \z[ \begin{smallmatrix}
  * h_3 & * h_2 \\
 h_4 & h_1
 \end{smallmatrix} \y](q) \\
   & \widetilde{\mc{V}}^{\frac{1}{2}}_{h} \z[ \begin{smallmatrix}
  * h_3 & * h_2 \\
 h_4 & h_1
 \end{smallmatrix} \y](q) = \sum_{r \in \mathbb{Z}_{\geq 0}+\frac{1}{2}} a_r q^{h+r-\frac{c}{24}}
\end{aligned}
\end{equation}
The coefficients $a_n$ should be negative when $h_1=h_4, ~ h_2=h_3$. The additional minus sign comes from exchanging the fermions:
\begin{equation}
	\widetilde{\mc{V}}^{\frac{1}{2}}_{h} \z[ \begin{smallmatrix}
  * h_3 & * h_2 \\
 h_4 & h_1
 \end{smallmatrix} \y](q) \ni  \langle V_1 \Lambda_2 | \psi_i \rangle \langle {\psi_i|\psi_i} \rangle^{-1} \langle \psi_i | \Lambda_3 V_4 \rangle,
\end{equation}
and note that when the internal state $\psi_i$ is fermionic, we have
\begin{equation}
	\langle V_1 \Lambda_2 | \psi_i \rangle  = - \langle \psi_i | \Lambda_2 V_1 \rangle ^*
\end{equation}
Now, let us compare the first few $a_r$'s. We denote the uncorrected one as $A_r$ and the corrected one as $a_r$. The leading order is
\begin{equation}
    \begin{aligned}
        A_{\frac{1}{2}} &= -2^{-1-c+4h} \frac{4(h^2+(h_1-h_2)^2)}{h} \\
        a_{\frac{1}{2}} &= -2^{-1-c+4h} \frac{4(h-h_1+h_2)^2}{h} 
    \end{aligned}
\end{equation}
Both $A_{\frac{1}{2}}$ and $a_{\frac{1}{2}}$ are non-positive. However, the corrected one $a_{\frac{1}{2}}$ has a double zero at $h=h_1-h_2$. This zero is indicated by the Ward identity
\begin{equation}
    \langle \Lambda_h(0) \Lambda_2(1) V_1(\infty) \rangle = -  \langle V_h(0) \p V_2(1) V_1(\infty) \rangle
\end{equation}
Thus, the correction is necessary even at the leading order. 

For higher orders, the formulae become complicated, so we do not present them here. When one assume $b,h_1,h_2>0$, both $A_{\frac{3}{2}}$ and $a_{\frac{3}{2}}$ are positive. At the third order, however, we find that $A_{\frac{5}{2}}$ is not always non-positive, while the corrected $a_{\frac{5}{2}}$ is indeed non-positive. Thus, we conclude that a correction is necessary for the elliptic block in \eqref{VHellip}.

\subsection{Crossing symmetry}
We can also check the crossing symmetry of the sphere correlation function of the Liouville theory. As in \eqref{fourpointfunc}, we can compute the crossing symmetries of the following correlation function
\begin{equation}
    \begin{aligned}
         & \left\langle V_{P_1}(0) W_{P_2}(z,\bar{z})W_{P_3}(1)V_{P_4}(\infty)\right\rangle_{g=0} \\
  &\qquad = \sum_P\Big(\widetilde{C}_b(P_4,P_3,P)\widetilde{C}_b(P,P_2,P_1) \mathcal{F}^1_{h_P}\left[
 \begin{smallmatrix}
  *h_{P_3} & *h_{P_2} \\
 h_{P_4} & h_{P_1}
 \end{smallmatrix}
  \right](z) \mathcal{F}^1_{h_P}\left[
 \begin{smallmatrix}
  *h_{P_3} & *h_{P_2} \\
 h_{P_4} & h_{P_1}
 \end{smallmatrix}
  \right](\bar z) \\
  & \qquad \qquad -C_b(P_4,P_3,P)C_b(P,P_2,P_1) \mathcal{F}^{\frac{1}{2}}_{h_P}\left[
 \begin{smallmatrix}
  *h_{P_3} & *h_{P_2} \\
 h_{P_4} & h_{P_1}
 \end{smallmatrix}
  \right](z) \mathcal{F}^{\frac{1}{2}}_{h_P}\left[
 \begin{smallmatrix}
  *h_{P_3} & *h_{P_2} \\
 h_{P_4} & h_{P_1}
 \end{smallmatrix}
  \right](\bar z) \Big)
    \end{aligned}
\end{equation}
Defining
\begin{equation}
    G_{1234}(z) = \langle O_1(0) O_2 (z,\bar z) O_3(1) O_4(\infty) \rangle
\end{equation}
Then the crossing equations are
\begin{equation}
    G_{1234}(z) = G_{3214}(1-z) = |z|^{2(h_4-h_3-h_3-h_1)}G_{1324}(z^{-1})
\end{equation}
For channel 3214, we need the Ward identity \cite{Du:2025lya}
\begin{equation}
    \langle WWVV\rangle=-\langle V\p\bp VVV\rangle+\langle V\p\bar{\Lambda}\bar{\Lambda}V\rangle+\langle V\bp\Lambda\Lambda V\rangle+\langle VWWV\rangle 
\end{equation}
where the ordering of the fields is fixed, and we only change the location of the supercharges. Note that the derivatives act on the operator with coordinate $(z,\bar z)$. 

Here, we would like to clarify the convergence property of $\sum_P$ of the conformal block expansion. The summations are over all the possible intermediate super-primaries, which should be understood as integrals rather than discrete sums:
\begin{equation}
 \sum_P\to -i\int_{i\mathbb{R}_{\geq 0}} dP \rho^{(b)}_{NS}(P)\z(...\y)
\end{equation}
There are two sources of divergence at large $h_P \simeq \frac{|P^2|}{2}$. One is the density 
\begin{equation}\label{partrho}
    \rho_{NS} \simeq e^{\pi Q |P|}
\end{equation}
and the other is the conformal block functions
\begin{equation}\label{partFF}
  \mathcal F_{h_P} \bar{\mathcal F}_{h_P} \simeq |16 q|^{|P|^2} 
\end{equation}
The conformal block expansion should converge in the unit disk of $q$. We aim to understand how this occurs for the (supersymmetric) Liouville theory. This requires determining the asymptotic behavior of the three-point function coefficient part $C_{43P}C_{P21}$ of the conformal block expansion. In \cite{alexanian2023barnes}, an approximation formula for the Barnes double gamma function $G(z,\tau)$ is introduced. Define the following functions
\begin{equation}
\begin{aligned}
& A(\tau) \equiv \frac{\tau}{2} \log (2 \pi \tau)+\frac{1}{2} \log (\tau)-\tau C(\tau) \\
& B(\tau) \equiv-\tau \log (\tau)-\tau^2 D(\tau)
\end{aligned}
\end{equation}
And
\begin{equation}
\begin{aligned}
C(\tau) & \equiv \sum_{k=1}^{m-1} \psi(k \tau)+\frac{1}{\tau} \log (\sqrt{2 \pi})-\sum_{\ell=0}^m \frac{B_{\ell} \tau^{\ell-1}}{\ell!} \psi^{(\ell-1)}(m \tau) \\
D(\tau) & \equiv \sum_{k=1}^{m-1} \psi^{(1)}(k \tau)-\sum_{\ell=0}^m \frac{B_{\ell} \tau^{\ell-1}}{\ell!} \psi^{(\ell)}(m \tau),
\end{aligned}
\end{equation}
where $\psi^{(m)}(z)$ are the polygamma functions, $B_\ell$ are the Bernoulli numbers. Here, m is a cutoff, which will be chosen as $m=10$. Define the polynomials $P_n(z ; \tau)$ as
\begin{equation}
\begin{aligned}
&P_n(z ; \tau)=\sum_{k=1}^n\binom{n+2}{k+2} \tau^{n-k} B_{n-k} \tilde{B}_{k+2}(z),\\
&\tilde{B}_k(z)=z^{-3}\left[B_k(z)-B_k(0)-B_k^{\prime}(0) z-B_k^{\prime \prime}(0) \frac{z^2}{2}\right], \quad k \geq 3
\end{aligned}
\end{equation}
where $B_k(z)$ are the Bernoulli polynomials. Define the following function
\begin{equation}
G_N(z ; \tau):=\frac{1}{\tau \Gamma(z)} e^{A(\tau) \frac{z}{\tau}+B(\tau) \frac{z^2}{2 \tau^2}} \prod_{m=1}^N \frac{\Gamma(m \tau)}{\Gamma(z+m \tau)} e^{z \psi(m \tau)+\frac{z^2}{2} \psi^{\prime}(m \tau)}
\end{equation}
where $N$ is another cutoff, which will be chosen as $N = 30$. Then, the approximation formula is for $z\in \mathbb C$ and $|\arg(\tau)|<\pi$,
\begin{equation}\label{approxG}
G(z ; \tau)=G_N(z ; \tau) \exp \left(z^3 \sum_{k=1}^M \frac{(-\tau)^{-k-1} P_k(z ;-\tau)}{k(k+1)(k+2)} \times N^{-k} + \mathcal O(N^{-M-1}) \right)
\end{equation}
where $M$ is a cutoff, which will be chosen as $M=10$. For $\operatorname{Re}(b)>0$, define
\begin{equation}
\Gamma_2 (z ; b )=(2 \pi)^{\frac{bz}{2}} b^{-\frac{z^2}{2}+\frac{z\left(b^{-1}+b\right)}{2 }-1} G\left(zb ; b^2\right)^{-1} 
\end{equation}
This function satisfies the same recursion relation as $\Gamma_b(z)$ in \eqref{shiftGamma}, but has a different normalization $\Gamma_2(1/b;b)=\sqrt{2\pi/b}$. Thus we have
\begin{equation}
    \Gamma_b(z) = \Gamma_2(z;b) f(b)
\end{equation}
In the three-point coefficient $C_b(p_i)$, the factors $f(b)$ will always be cancelled. Practically, it will be more stable to compute the logarithm of the three-point coefficient. Note that $P_k(z;\tau)$ is a monic polynomial of degree $(n-1)$ in $z$. Thus $\ln G(z;\tau)$ contains $z^{12}$ for $M=10$, and the coefficient will be negative if the other parameters are properly chosen\footnote{This means that the central charge and the conformal weights of the external states are positive. }. This will give rise to an $e^{-\# P^{12}}$ suppression for the three-point function coefficients, and thus make the $\sum_P$ integral in the conformal block expansion highly convergent. Actually, the asymptotic expansion in the $z\to \infty$ limit is
\begin{equation}
    \begin{aligned}
        \ln G(z ; \tau) & =\left(a_2(\tau) z^2+a_1(\tau) z+a_0(\tau)\right) \ln (z) \\
& \qquad +b_2(\tau) z^2+b_1(\tau) z+b_0(\tau)+\mathcal{E}(z ; \tau), \\
    \end{aligned}
\end{equation}
where $\mathcal E (z;\tau) = \mathcal O(z^{-1})$, and
\begin{equation}
\begin{aligned}
 a_0(\tau)&=\frac{\tau}{12}+\frac{1}{4}+\frac{1}{12 \tau}, \\
 a_1(\tau)&=-\frac{1}{2}\left(1+\frac{1}{\tau}\right), \\
 a_2(\tau)&=\frac{1}{2 \tau}, \\
 b_1(\tau)&=\frac{1}{2}\left(\left(\frac{1}{\tau}+1\right)(1+\ln (\tau))+\ln (2 \pi)\right), \\
 b_2(\tau)&=-\frac{1}{2 \tau}\left(\frac{3}{2}+\ln (\tau)\right),
\end{aligned}
\end{equation}
Using this formula, one can show that
\begin{equation}
    \ln C_b(P_1,P_2, iP) = -2 P^2\ln 2 - \frac{\pi Q}{2} P + \mathcal O(1),\quad P \to \infty  
\end{equation}
And similarly for $\widetilde{C}_b(P_1,P_2,iP)$. Thus, the factor $C_{43P}C_{P21}$ will contribute
\begin{equation}\label{3.30}
    C_{43P}C_{P21} \simeq 16^{-|P|^2} e^{-\pi Q|P|}
\end{equation}
Combining with \eqref{partrho} and \eqref{partFF}, we understand that as long as $|q|<1$, the conformal block expansion converges rapidly with a suppression of the form $e^{-\# |P|^2}$. In \cite{Das:2020uax,Collier:2019weq}, similar leading order behavior of the light-light-heavy three-point function coefficients in compact CFTs has been observed\footnote{We thank Sridip Pal for helpful clarification regarding the asymptotic behaviors of 2D CFTs. }, although Liouville theory is non-compact. Moreover, the equation \eqref{3.30} can be stated in a more general form
\begin{equation}
   16^{h+\bar h} \times C_{43h}C_{h21} \simeq e^{-\frac{S(h,\bar h)}{2}}
\end{equation}
where $S(h,\bar h)$ is the entropy function. 

Since the convergence property is good, in real computation, we can trust the approximation formula \eqref{approxG} which makes the integral converge even faster. We test the crossing symmetry for various data\footnote{We thank Zhengyuan Du for helping with the coding and debugging. }. For instance, we take $b=\frac{e}{\pi}$ and $p_1 = \frac{1}{3},~p_2=\frac{1}{2},~p_3=\frac{1}{7},~p_4=\frac{1}{4}$ so that
\begin{equation}
    h_i = \frac{Q^2}{8} - \frac{p_i^2}{2},\quad i=1,2,3,4
\end{equation}
The location of the second operator is chosen to be $z=1+0.1\times j,~j=1,2,\cd ,9$. The results are shown in Figure \ref{fig:main}. Before the correction, there is a visible error between different channels. After the correction, the error is less than order $10^{-3}\%$. 
\begin{figure}[H]
 \centering
 \begin{subfigure}{\textwidth}
  \centering
  \includegraphics[width=0.8\textwidth, height=8cm, keepaspectratio]{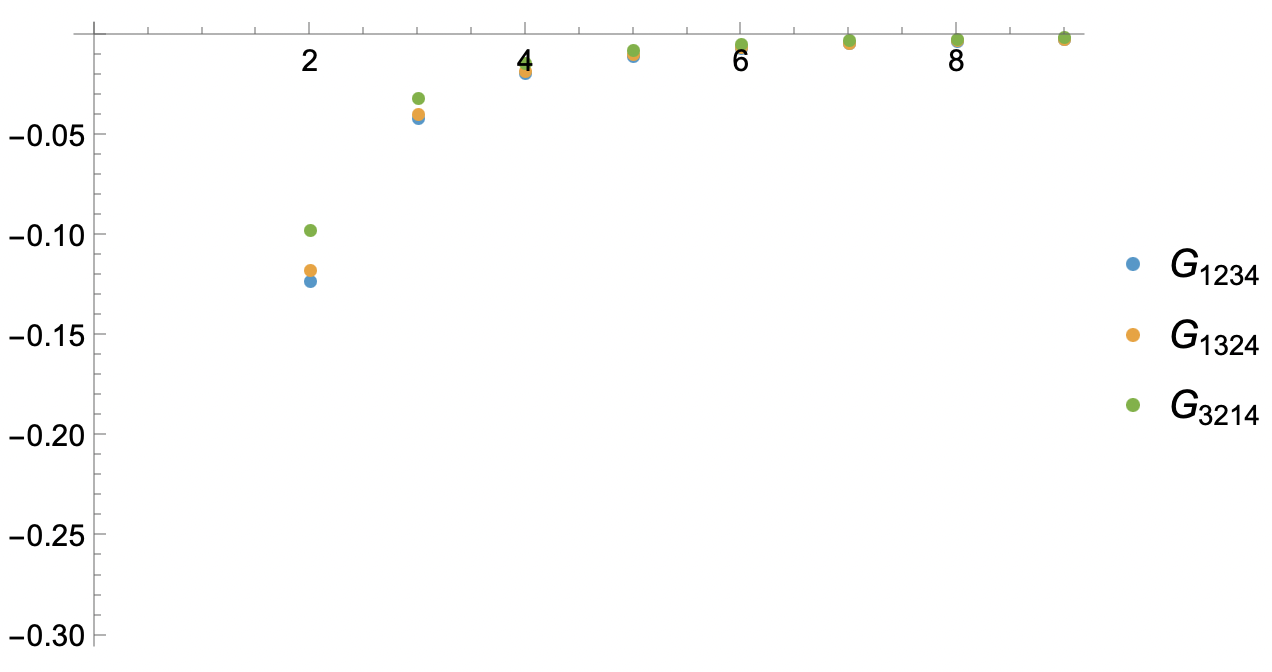}
  \caption{Uncorrected blocks: The error is about $10\%$.}
  \label{fig:sub1}
 \end{subfigure}
 \label{fig:main}
\end{figure}

\begin{figure}[H]
 \ContinuedFloat  
 \centering
 \begin{subfigure}{\textwidth}
  \centering
  \includegraphics[width=0.8\textwidth, height=8cm, keepaspectratio]{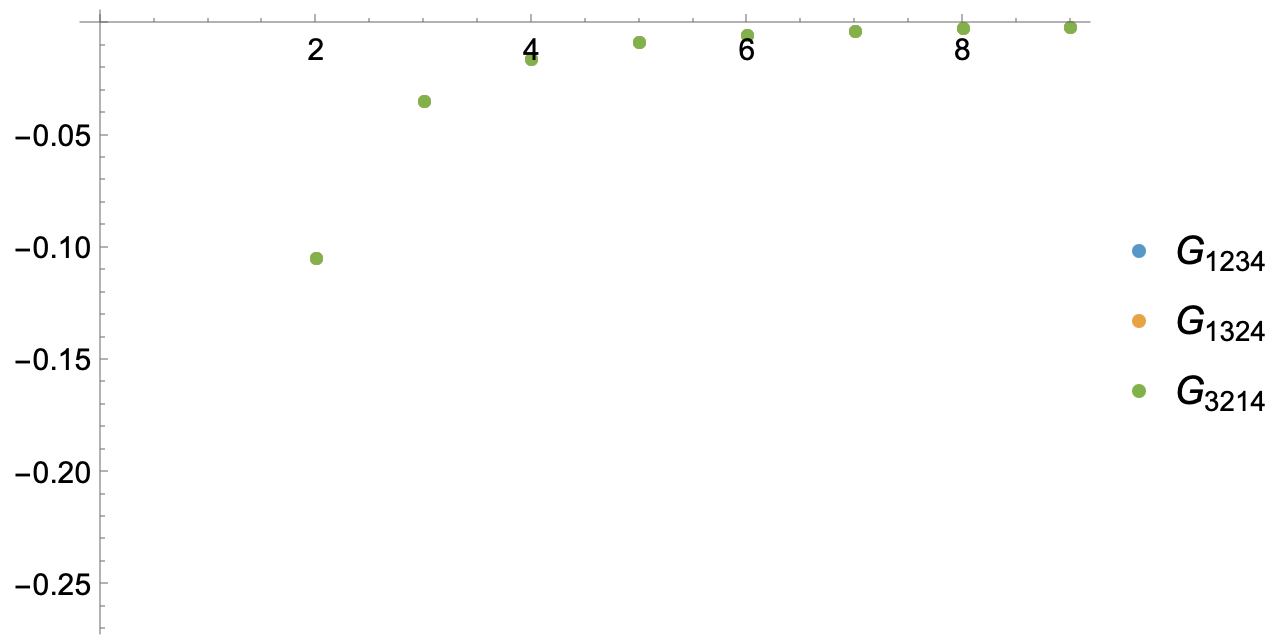}
  \caption{Corrected blocks: The error is less than $10^{-3}\%$.}
  \label{fig:sub2}
 \end{subfigure}
 \caption{Crossing symmetry of $ \left\langle V_{P_1}(0) W_{P_2}(z,\bar{z})W_{P_3}(1)V_{P_4}(\infty)\right\rangle_{g=0}$}  
 \label{fig:main}
\end{figure}

\subsection{Direct checks}
For higher orders of $q$, say $q^{\frac{9}{2}}$ and beyond, the crossing symmetries may not be sensitive enough to check the formula. We therefore adopt the most direct verification: first compute the $\mathcal{F}_h^{\frac{1}{2}}$ block via $c$-recursion, then use \eqref{FandH} to obtain $\mathcal{H}_h^{\frac{1}{2}}$, and finally compare with the result from the $h$-recursion \eqref{Hrec}. 

We compare each order of the $q$-expansion. For $b=\frac{e}{\pi}$ and $p_1 = \frac{1}{3},~p_2=\frac{1}{2},~p_3=\frac{1}{7},~p_4=\frac{1}{4}$, we set $h=\frac{j}{5}+10^{-4}$ with $j=0,1,2,\cd,9$. The results of the two methods agree with each other with an error below $10^{-4}\%$. Note that the norm of the $h_i$'s is about $0.25$, and the choice of the norm of $h$ is basically of the same order or a difference of an order of magnitude. If the magnitude of $h$ is far away from the external conformal weights $h_i$, say a difference of three orders of magnitude, there could be a larger error between the two methods, which is common in numerical computation. However, as we have analyzed in the previous subsection, when $|q|<1$, the contribution of heavy internal states is suppressed by the factor $e^{-\# |P|^2}$ in the integrand. Thus, the two methods should stably give consistent results. Indeed, the author has also checked the crossing symmetries for various four-point functions using both methods, and they matched very well. 

We can also test the cases with complex parameters. For $b=\frac{e}{\pi}e^{i\frac{\pi}{4}}$ and $p_1 = \frac{1}{3}e^{-i\frac{\pi}{4}},~p_2=\frac{1}{5}e^{-i\frac{\pi}{4}},~p_3=\frac{1}{7}e^{-i\frac{\pi}{4}},~p_4=\frac{1}{4}e^{-i\frac{\pi}{4}}$, we set $h=\frac{j}{5}+10^{-4}$ with $j=0,1,2,\cd,9$. Again, the error is less than order $10^{-4}\%$.

\section{Conclusion}\label{sec4}
In this work, we have completed the construction of the elliptic superconformal blocks for the NS sector of \(\mathcal N=1\) super Liouville theory by determining the regular part of the \(\frac12\)-block with two starred external fields up to order \(\mathcal O(q^{15/2})\). The correction term, given in Eq.~\eqref{correctionfunc}, is a polynomial of degree two in the external conformal weights with coefficients depending rationally on the Liouville parameter \(b\) and satisfying the expected self-duality and permutation symmetries.

The necessity of this correction was demonstrated explicitly through several numerical tests. In the pillow geometry, the uncorrected blocks can lead to positive coefficients that should be negative due to the positivity of the norms and fermion exchange, while the corrected ones have the correct zero indicated by the Ward identity at the leading order, as well as the correct sign at subleading orders. Crossing symmetry of the \(\langle VW WV\rangle\) four-point function, which was violated at the \(10\%\) level without the correction, is restored to an accuracy better than \(10^{-3}\%\) after including our term. Direct comparison of the elliptic blocks obtained from the \(c\)-recursion and the \(h\)-recursion confirms the correctness of the expansion coefficients up to order \(\mathcal O(q^{15/2})\) with an error below \(10^{-4}\%\) for a wide range of parameters, including complex values of \(b\) and the external momenta.

The availability of a reliable and efficient \(h\)-recursion formula for all types of NS-sector superconformal blocks opens the way to high-precision and more efficient numerical bootstrap studies of \(\mathcal N=1\) super Liouville theory and, more generally, of two-dimensional supersymmetric CFTs and 2d superstrings. 

Finally, let us comment on aspects that are not covered in this work. Firstly, one may worry about whether similar corrections are needed in the Ramond sector. At first glance, there should be no similar problem in the R sector, as the regular parts of the elliptic blocks are all of order $\mathcal O(h^0)$, see \cite{Suchanek:2010kq}. More importantly, as we cannot prove the proposal \eqref{mainprop} at this stage, a deeper understanding of the $\mathcal O(1/h)$ correction of the classical blocks $\ln \mathcal F$ of the $\mathcal N=1$ SCFTs is desirable. 

\section*{Ackowledgements }
The author would like to thank Chi-Ming Chang, Zhengyuan Du, Lorenz Eberhardt, Sridip Pal, Victor A. Rodriguez, and Zhe-fei Yu for useful discussions. K.L. is supported by the NSFC special fund for theoretical physics No. 12447108 and the National Key Research and Development Program of China No. 2020YFA0713000. K.L also thanks the Yukawa Institute for Theoretical Physics at Kyoto University, and the organizers of the workshop “Progress of Theoretical Bootstrap” for their hospitality during the course of this work.

\bibliographystyle{JHEP}
\bibliography{ref.bib}

\end{document}